\documentclass[useAMS,usenatbib]{mn2e}
\usepackage{aas_macros,graphicx,times,multirow}
\title[X-ray and optical observations of \src]
  {Early X-ray and optical observations of the soft gamma-ray repeater SGR\,0418+5729}
\author[P. Esposito et al.]
{P.~Esposito,$^{1,2}$\thanks{E-mail: paoloesp@iasf-milano.inaf.it} G.~L.~Israel,$^{3}$ R.~Turolla,$^{4,5}$ A.~Tiengo,$^{1}$ D.~G\"otz,$^{6}$ A.~De~Luca,$^{1,7}$
\newauthor R.~P.~Mignani,$^{5}$ S.~Zane,$^{5}$ N.~Rea,$^{8}$ V.~Testa,$^{3}$ P.~A.~Caraveo,$^{1}$ S.~Chaty,$^{6}$ F.~Mattana,$^{9}$ %P.~G.~Jonker,$^{9,10}$ 
 \newauthor   S.~Mereghetti,$^{1}$ A.~Pellizzoni$^{10}$ and P.~Romano$^{11}$
\smallskip\\
$^1$INAF/Istituto di Astrofisica Spaziale e Fisica Cosmica - Milano, via E.~Bassini 15, I-20133 Milano, Italy\\
$^2$INFN - Istituto Nazionale di Fisica Nucleare, sezione di Pavia, via A.~Bassi 6, I-27100 Pavia, Italy\\
$^3$INAF/Osservatorio Astronomico di Roma, via Frascati 33, I-00040 Monteporzio Catone, Italy\\
$^4$Universit\`a di Padova, Dipartimento di Fisica, via F.~Marzolo 8, I-35131 Padova, Italy\\
$^5$Mullard Space Science Laboratory, University College London, Holmbury St. Mary, Dorking, Surrey RH5 6NT, UK\\
$^6$AIM (UMR 7158 CEA/DSM-CNRS-Universit\'e Paris Diderot) Irfu/Service d'Astrophysique, Saclay,  FR-91191 Gif-sur-Yvette Cedex, France\\
$^7$IUSS - Istituto Universitario di Studi Superiori, viale Lungo Ticino Sforza 56, I-27100 Pavia, Italy\\
$^8$Institut de Ci\`encies de l'Espai (CSIC--IEEC), Campus UAB, Facultat de Ci\`encies, Torre C5-parell, E-08193 Barcelona, Spain\\
$^{9}$Laboratoire AstroParticule et Cosmologie, Universit\'e Paris 7 - Denis Diderot, rue A.~Domon et L.~Duquet 10, FR-75205 Paris, France\\
$^{10}$INAF/Osservatorio Astronomico di Cagliari, localit\`a Poggio dei Pini, strada 54, I-09012 Capoterra, Italy\\
$^{11}$INAF/Istituto di Astrofisica Spaziale e Fisica Cosmica - Palermo, via U.~La Malfa 153, I-90146 Palermo, Italy}
\date{Accepted 2010 February 18. Received 2010 February 10; in original form 2010 January 04}
\pagerange{\pageref{firstpage}--\pageref{lastpage}} \pubyear{2009}
\def\LaTeX{L\kern-.36em\raise.3ex\hbox{a}\kern-.15em
    T\kern-.1667em\lower.7ex\hbox{E}\kern-.125emX}
\def\xmm {\emph{XMM-Newton}}
\def\cxo {\emph{Chandra}}
\def\swift {\emph{Swift}}

\def\xte {\emph{RXTE}}
\def\rst {\emph{ROSAT}}

\def\src {SGR\,0418+5729}
\def\flux {\mbox{erg cm$^{-2}$ s$^{-1}$}}
\def\lum {\mbox{erg s$^{-1}$}}
\def\nh {$N_{\rm H}$ }

\begin{document}
%\special{!userdict begin /bop-hook{gsave 150 90 translate 55 rotate /Times-Roman findfont 150 scalefont setfont 0 0 moveto 0.9 setgray (Confidential) show grestore}def end}
\label{firstpage}
\maketitle
\begin{abstract}
Emission of two short hard X-ray bursts on 2009 June 5 disclosed the existence of a new soft gamma-ray repeater, now catalogued as \src. After a few days, X-ray pulsations at a period of 9.1 s were discovered in its persistent emission. \src\ was monitored almost since its discovery with the \emph{Rossi X-ray Timing Explorer} (2--10 keV energy range) and observed many times with \swift\ (0.2--10 keV). The source persistent X-ray emission faded by a factor $\sim$10 in about 160 days, with a steepening in the decay about 19 days after the activation. The X-ray spectrum is well described by a simple absorbed blackbody, with a temperature decreasing in time. A phase-coherent timing solution over the $\sim$160 day time span yielded no evidence for any significant evolution of the spin period, implying a 3$\sigma$ upper limit of $1.1\times10^{-13}$ s s$^{-1}$ on the period derivative and of $\sim$$3\times10^{13}$ G on the surface dipole magnetic field. Phase-resolved spectroscopy provided evidence for a significant variation of the spectrum as a function of the stellar rotation, pointing to the presence of two emitting caps, one of which became hotter during the outburst. Finally, a deep observation of the field of \src\ with the new Gran Telescopio Canarias 10.4-m telescope allowed us to set an upper limit on the source optical flux of $i'>25.1$ mag, corresponding to an X-ray-to-optical flux ratio exceeding $10^4$, consistent with the characteristics of other magnetars.
\end{abstract}

\section{Introduction}
Soft gamma-ray repeaters (SGRs) form a small class of sources which emit short ($\sim$0.1 s) bursts of soft gamma-rays with peak luminosities of $\sim$$10^{41}$--$10^{42}$ erg~s$^{-1}$. 
%They were originally thought to be a peculiar subclass of gamma-ray bursts (GRBs) but, contrary to the behaviour of classical GRBs, SGRs produce series of bursts over various time scales and the events are characterised by a soft spectral shape. 
Originally believed to be a subclass of the gamma-ray bursts (GRBs), they were soon recognised as a different population because their bursts repeat and are quite soft, at variance with canonical GRBs.
Apart from the flaring activity, sometimes culminating in dramatic, very energetic \emph{giant flares} with peak luminosity exceeding $10^{46}$ erg s$^{-1}$ \citep[as it was the case of the 2004 event from SGR\,1806--20;][]{hurley05short}, SGRs display persistent emission from 0.1 to hundreds of keV, with pulsations of several seconds ($P\sim2$--9 s), rapid spin down ($\dot{P}\sim10^{-11}$ s s$^{-1}$), and with significant variability in period derivative, pulse shape, flux, and spectral shape.\\
\indent This phenomenology has been interpreted within the context of the \emph{magnetar} model \citep{paczynski92,duncan92,thompson95,thompson96}. SGRs are in fact believed to be young, isolated neutron stars endowed with an ultrahigh magnetic field (\mbox{$B\sim10^{14}$--$10^{15}$ G}), which is thought to be the ultimate energy reservoir for their activity. It is now commonly accepted that anomalous X-ray pulsars (AXPs), a peculiar class of pulsating X-ray sources, are also magnetars, in view of the many similarities in their persistent and bursting emission (see \citealt{woods06} and \citealt{mereghetti08} for recent reviews).\footnote{Ten AXPs and six SGRs are confirmed, and there are a few candidates; see catalog at \mbox{http://www.physics.mcgill.ca/$\sim$pulsar/magnetar/main.html}.}\\
\indent On 2009 June 5, a new SGR, namely \src, was discovered thanks to the detection of two short X-ray bursts with the \emph{Fermi} Gamma-ray Burst Monitor (GBM) and other hard X-ray instruments \citep{vanderhorst09short,vanderhorst10short}. After a few days, on 2009 June 10, X-ray pulsations at a period of 9.1 s were discovered in the persistent emission with \emph{Rossi X-ray Timing Explorer} (\xte), further supporting the magnetar nature of the source \citep*{gogus_atel2076_09}. Ground-based follow-up observations in optical and infrared wavebands (discussed in the following), as well as at radio wavelengths (with the 100-m Green Bank Telescope; \citealt{lorimer09}), failed to detect the source.\\
\indent Here we report on the analysis of a series of observations carried out in the soft X-ray range (1--10 keV) by the X-Ray Telescope aboard \swift\ and by the Proportional Counter Array instrument of the \xte\ satellite, collected in the period from 2009 June 11 to 2009 November 24. We also present the results of a very deep optical observation performed on 2009 September 15 with the new Gran Telescopio Canarias 10.4-m telescope on La Palma (Canary Islands, Spain), which is presently the largest single-aperture optical telescope in the world.

\section{Observations and data reduction}
\subsection{\swift}
The X-Ray Telescope (XRT; \citealt{burrows05short}) on-board \swift\ uses a CCD detector sensitive to photons between 0.2 and 10 keV with an effective area of about 110 cm$^2$ and a field of view (fov) of 23$'$ diameter. Several observations of \src\ were performed following its discovery, in both photon counting (PC) and windowed timing (WT) modes. In PC mode the entire CCD is read every 2.507 s, while in WT mode only the central 200 columns are read and only one-dimensional imaging is preserved, achieving a time resolution of 1.766 ms (see \citealt{hill04short} for more details). Table~\ref{xrt-log} (see also Fig.~\ref{xte_decay}) reports the log of the observations that were used for this work.\\
\indent The data were processed with \textsc{xrtpipeline} (version 0.12.3, in the \textsc{heasoft} software package\footnote{See http://heasarc.gsfc.nasa.gov/docs/software/ftools/ftools\_menu.html.} 6.6), and standard filtering and screening criteria were applied by using \textsc{ftools} tasks.We extracted the PC source events from a circle with a radius of 20 pixels (one pixel corresponds to about $2\farcs36$) and the WT data from a $40 \times40$ pixels box along the image strip. To estimate the background, we extracted PC and WT events from source-free regions far from the position of \src.\\
\indent For the timing analysis, the photon arrival times were converted to the Solar system barycentre with the \textsc{barycorr} task and using the \cxo\ source position ($\rm RA =04^h18^m33\fs867$, $\rm Dec. =+57\degr32'22\farcs91$, epoch J2000; $0\farcs35$ accurate) provided by \citet{woods09}. For the spectral fitting, the ancillary response files (arf) were generated with \textsc{xrtmkarf}, and they account for different extraction regions, vignetting and point-spread function corrections. We used the latest available spectral redistribution matrix (rmf) in \textsc{caldb} (v011).
\begin{table}
\centering
%\begin{minipage}{10.3cm}
\caption{Journal of the \swift/XRT observations of \src. The complete observation sequence number is prefixed by 00031422 followed by the three digit segment number given here (e.g. 00031422001).}
\label{xrt-log}
\begin{tabular}{@{}lccc}
\hline
Sequence/mode & \multicolumn{2}{c}{Start/End time (UT)} & Exposure$^{a}$\\
 & \multicolumn{2}{c}{yyyy-mm-dd hh:mm} & (ks)\\
\hline
001 / PC & 2009-07-08~20:52 & 2009-07-08~22:54 & 2.9\\
002 / PC & 2009-07-09~00:07 & 2009-07-09~19:29 & 10.6\\
003 / PC & 2009-07-10~00:17 & 2009-07-10~07:01 & 5.6\\
004 / WT & 2009-07-12~00:29 & 2009-07-12~15:04 & 7.1\\
006 / WT & 2009-07-15~15:16 & 2009-07-15~23:40 & 7.7\\
007 / WT & 2009-07-16~00:55 & 2009-07-16~23:36 & 16.4\\
008 / PC & 2009-09-20~21:11 & 2009-09-21~23:04 & 9.4\\
009 / PC & 2009-09-22~00:46 & 2009-09-22~23:28 & 7.6\\
010 / PC & 2009-11-08~00:38 & 2009-11-08~23:03 & 15.1\\
\hline
\end{tabular}
\begin{list}{}{}
\item[$^{a}$] The exposure time is usually spread over several snapshots (single continuous pointings of the target) during each observation.
\end{list}
%\end{minipage}
\end{table}

\subsection{\xte}
The Proportional Counter Array (PCA; \citealt{jahoda96}) on-board \xte\ consists of an array of five collimated xenon/methane multianode Proportional Counter Units (PCUs) operating in the 2--60 keV energy range, with a total effective area of approximately 6500 cm$^2$ and a full width at half maximum (fwhm) fov of about 1$\degr$. The 46 \xte/PCA observations of \src\ (obs. ID: 94048) reported here span from 2009 June 11 to 2009 November 24 (see Fig.~\ref{xte_decay}). The exposures range from 0.6 ks to 13.6 ks, for a total of 194.2 ks.\\
%\begin{figure}
%\resizebox{\hsize}{!}{\includegraphics[angle=0]{observations.jpg}}
%\caption{\label{observations} Epochs of observations of \src\ with \xte\ (in black), \swift\ (in red), and the OSIRIS camera of the Gran Telescopio Canarias (in blue).}
%\end{figure}
\indent Raw data were reduced using the \textsc{ftools} package. To study the timing properties of \src, we restricted our analysis to the data in Good Xenon mode, with a time resolution of 1 $\mu$s and 256 energy bins. After correction to the Solar system barycentre with the \textsc{fxbary} task, the event-mode data were extracted in the 2--10 keV energy range from all active detectors (in a given observation) and all layers, and binned into light curves of 0.1 s resolution.\\
\indent For the spectral analysis, we have only made use of the Standard 2 data (time resolution of 16 s and 128 energy bins) from PCU-2, which is the best-calibrated proportional counter unit in the PCA, and spectra were extracted including only top-layer events. Good time intervals were selected to ensure stable pointing and filter out electron contamination events, following the standard procedure for faint sources. Response matrices (combining rmf and arf files) were generated using \textsc{pcarsp}. Background subtraction for PCA data is performed using a synthetic model that accounts for the time-variable particle background and for the average cosmic X-ray background (see \citealt{jahoda06} for details). Possible Galactic diffuse emission remains in the background-subtracted data, as well as emission from any X-ray source that occurs in the instrument fov.

\section{Spectral analysis}
Because of its nonimaging nature and wide field of view, the \xte/PCA instrument is not well suited for the spectral study of relatively faint and soft sources like \src. The \swift/XRT instrument, instead, offers a good sensitivity and ensures proper background subtraction (also in the case of WT observations). For this reason we started the spectral analysis by working only with the \swift/XRT data.
\label{swiftspec}
\begin{table*}
\centering
\begin{minipage}{10.5cm}
\caption{\swift\ spectral results for \src\ for the single blackbody model. Errors are at 1$\sigma$ confidence level for a single parameter of interest. The \nh\ value is $1.1^{+0.1}_{-0.2}\times10^{21}$ cm$^{-2}$.}
\label{xrt-spec}
\begin{tabular}{@{}lcccc}
\hline
Sequence & Time since $t_0$$^{a}$ & $kT$ & Radius$^{b}$ & Absorbed flux$^{c}$\\
 & (days) & (keV) & (km) & ($10^{-11}$ \flux)\\
\hline
001--003& 33.004--34.430 &$0.96\pm0.01$ &$0.62\pm0.02$ & $1.31\pm0.03$\\
004--007& 36.156--41.121 &$0.94\pm0.01$ &$0.65\pm0.02$ & $1.35\pm0.02$\\
008--009& 107.020--109.113 &$0.84\pm0.02$ &$0.43\pm0.02$ & $0.37\pm0.01$\\
010     & 155.164--156.098 &$0.83\pm0.03$ &$0.33\pm0.02$ & $0.21\pm0.01$\\
\hline
\end{tabular}
\begin{list}{}{}
\item[$^{a}$] We assumed as $t_0$ MJD 54987.862, which is the time of the detection of the first burst.
\item[$^{b}$] The blackbody radius is calculated at infinity and for a distance of 5 kpc.
\item[$^{c}$] In the 1--10 keV energy range.
\end{list}
\end{minipage}
\end{table*}
\subsection{\swift/XRT}
The observations carried out within a few days and in which the count rates were similar were added in order to accumulate sufficient statistics for meaningful spectral fits; as a result we obtained four spectra: 001--003, 004--007 (WT data), 008--009, and 010 (see Table~\ref{xrt-log}). The spectral channels were then grouped to have bins with a minimum number of 30 photons. The spectral fitting, with the \textsc{xspec} analysis package (version 12.4; \citealt{arnaud96}), was performed in the 0.7--10 keV energy range, since the very few counts from \src\ made the data below 0.7 keV useless.\\
\indent We first tried fitting simple models to the spectra: a power law, a blackbody, a double blackbody, and a blackbody plus power law, all modified for interstellar absorption. The abundances used are those of \citet{anders89} and photoelectric absorption cross-sections are from \citet{balucinska92}. We fit all spectra simultaneously with the hydrogen column density tied up between the four datasets. While the power-law model is disfavoured ($\chi_\nu^2=2.10$ for 384 degrees of freedom [dof]), the single blackbody provides an acceptable fit ($\chi_\nu^2=1.16$ for 384 dof); see Table~\ref{xrt-spec} for the best-fitting parameters. Both double blackbody and blackbody plus power law yield similarly good fits ($\chi^2_\nu$ values of 1.10 for 376 dof and 1.12 for 376 dof, respectively), but the parameters of the additional spectral components are rather unconstrained. Therefore we prefer the simplest model, a single blackbody with temperature $kT$ changing from $\sim$1 keV to $\sim$0.8 keV (the derived absorbing column is $N_{\rm H}\simeq1.1\times10^{21}$ cm$^{-2}$). The corresponding observed flux decreases from $\sim$$1.3\times10^{-11}$ \flux\ to $\sim$$2\times10^{-12}$ \flux\ (1--10 keV). These results are consistent with those reported by \citet{cummings09} and \citet*{gogus_atel2121_09}, which were obtained using a smaller dataset.\\
\indent Although the present spectra do not allow us to discriminate among more complex (e.g. two component or more physical) spectral models, magnetar spectra in the $\sim$1--10 keV range are typically described by a combination of a soft, thermal component (a blackbody) and a harder one. The latter is often represented by a non-thermal power-law tail, although in some sources a second, hotter blackbody provides an equally good (or better) interpretation (e.g. \citealt*{gotthelf05,tiengo08}).
In the magnetar model, the power-law tail naturally arises from resonant cyclotron scattering (RCS) of seed thermal photons on magnetospheric currents \citep*{tlk02}. In the light of this, we attempted a simultaneous fit of all spectra with two different RCS models: the one-dimensional (1D) model by \citet{lyutikov06} (see also \citealt{rea08}) and the NTZ model, based on three-dimensional Monte Carlo simulations (\citealt*{nobili08}; \citealt{zane09}).\\
\indent We tried several fits, first leaving all parameters free (except for the column density), then assuming that either the surface temperature or some of the magnetospheric parameters are the same in all observations (the optical depth in the 1D model and the twist angle in the NTZ model). In all cases, although some fluctuations are allowed in the magnetospheric properties, we found that the best fits ($\chi_\nu^2 = 1.15$ for both models) are obtained by invoking scenarios in which the major change among the different observations is in terms of a decrease in the temperature of the thermal seed photons (from $\sim$0.9 keV to $\sim$0.75 keV in the 1D model and from $\sim$1 keV to $\sim$0.9 keV in the NTZ fits) accompanied by a decrease in the emitting surface area (of a factor $\sim$4 and $\sim$7 in RCS and NTZ, respectively). The reprocessing of the thermal radiation by the magnetosphere is always quite modest and these results are consistent with the single blackbody fits discussed above. Overall, this favours a scenario in which heat is deposited in a spot on the neutron-star surface, following the bursting activity, that cools off and shrinks as time elapses.

\subsection{\xte/PCA}\label{rxtespec}
\begin{figure}
\resizebox{\hsize}{!}{\includegraphics[angle=-90]{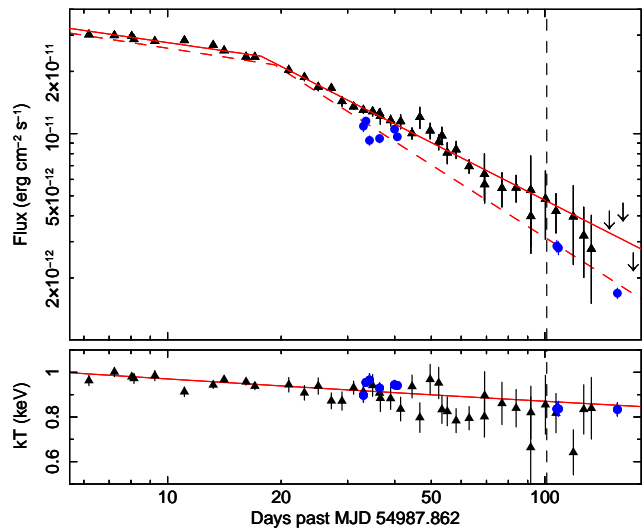}}
\caption{\label{xte_decay} Top panel: the outburst decay of the persistent X-ray flux of \src\ as seen by \xte\ (black filled triangles) and \swift\ (blue filled circles). We assumed as $t=0$ MJD 54987.862, which is the time of the detection of the first burst. The fluxes are absorbed and in the 2--10 keV energy range; the down-arrows indicate upper limits at 3$\sigma$ confidence level. The broken power law describing the \xte\ flux decay is superimposed in red (solid line); we also plotted (red, dashed line) the simultaneous fit of the \swift\ and \xte\ data after subtracting a constant flux from the latter (see Section~\ref{rxtespec}). Bottom panel: evolution of the characteristic temperature of the blackbody model inferred from the spectral fitting. The power-law model describing the decay of the blackbody temperature (see Section~\ref{disc}) is plotted in red. The vertical dashed lines indicate the epoch of the observation with the Gran Telescopio Canarias (see Section~\ref{grantecan}).}
\end{figure}

The analysis of the PCA spectra was mainly aimed at having flux measurements over the entire data span. We thus adopted the absorbed blackbody model and fitted it to all the PCA spectra together; the parameters were left free to vary, except for the absorption column density that was fixed at the value measured with \swift. Spectral channels having energies below 3 keV (standard for \xte\ spectral analysis -- see, e.g., \citealt{park04short}) and above 10 keV, owing to the very low signal-to-noise ratio, were ignored. This resulted in an acceptable fit ($\chi^2_\nu=0.78$ for 644 dof) with spectral parameters similar to those reported in Table~\ref{xrt-spec} (see the bottom panel of Fig.~\ref{xte_decay}, but see also below for indication of a likely background contamination in these data). Again, there is a clear indication of an overall spectral softening during the decay.\\
\indent We plot the resulting long-term light curve in the top panel of Fig.~\ref{xte_decay} (the last three observations provided only upper limits). The flux decays as a broken power law. Fixing $t=0$ at the time of the detection of the first burst (2009 June 5 at 20:40:48.883 UTC; \citealt{vanderhorst10short}), the break occurs at $17.7\pm0.7$ days, when the index changes from $\alpha_1=-0.27 \pm 0.02$ to $\alpha_2=-0.93 \pm 0.04$ (here and in the following all uncertainties are at 1$\sigma$ confidence level). It is important to note that the fit parameters depend on the assumed time origin, as well as on a possible steady component of the SGR flux. The uncertainties introduced by making different (reasonable) assumptions are larger than the statistical errors reported here by a factor $\sim$5.\\
\indent A comparison with the \swift\ flux measurements indicates a substantial discrepancy between the values derived from the XRT and the PCA instruments. This is apparent in Fig.~\ref{xte_decay} where we plotted the results from the individual \swift/XRT observations. While the uncertainty in the relative calibration of the two instruments (which is 10--20\% in the energy band considered here; \citealt{kirsch05short}) cannot account for the mismatch, the fact that the difference between the \swift/XRT fluxes and the best-fitting curve of the \xte\ data is consistent with being constant ($\Delta F = (1.7\pm0.1)\times10^{-12}$ \flux) suggests the presence of a background contribution to the \xte\ data. After subtracting $\Delta F$ from the \xte\ fluxes, we fit a broken power law to the combined \xte\ and \swift\ data, obtaining slightly different parameters for the decay: $\alpha'_1=-0.28 \pm 0.02$, $\alpha'_2=-1.17 \pm 0.02$, and $19.2\pm0.4$ days for the break epoch.\\
%rosat: 0.0971300 (0.0629910-0.131269); swift: 0.0056415 (0.0043915-0.0068915)
\indent With respect to the flux discrepancy, we note that it is consistent with the fluctuations of the cosmic X-ray background, which in the PCA produce effects of the order of $4\times10^{-12}$ \flux\ in the 2--10 keV band \citep{jahoda06}. We tried to better assess the situation by examining the X-ray images  of the field. Apart from \src, eight point sources are detected with a signal-to-noise above 3 in the 2--10 keV band within the 23$'$-diameter fov of the \swift/XRT detector. Assuming for the absorption the total Galactic value in the direction of \src\ ($\sim$$5.6\times10^{21}$ cm$^{-2}$, see Section~\ref{disc}), we estimate that their total contribution to the flux observed in \xte\ for a blackbody spectrum with $kT=0.2$ keV is $\approx$(1.5--$2.5)\times10^{-13}$ \flux, $\approx$(4--$6)\times10^{-13}$ \flux\ for a power-law spectrum with photon index $\Gamma =2$. Four sources outside the XRT fov, but still in the PCA one, are present in the \rst\ All-Sky Survey Faint Source Catalog (\citealt{voges00short}; see also Section~\ref{rosat}); using the same assumptions as above, their 2--10 keV flux is estimated to be of $\approx$(3--$6)\times10^{-14}$ \flux\ in the case of the blackbody spectrum, and $\approx$(2--$5)\times10^{-12}$ \flux\ for the power-law spectrum.
\section{Timing analysis}\label{timing}
\begin{figure}
\resizebox{\hsize}{!}{\includegraphics[angle=-90]{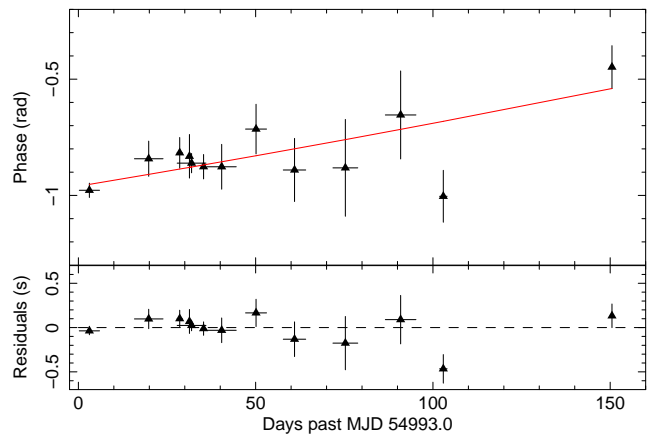}}
\caption{\label{residual} Top panel: 2--10\,keV \xte/PCA (binned at 10 days) and \swift/XRT pulse phase evolution with respect to the period measured by \citet*{gogus_atel2076_09} as a function of time. The red solid line represents the phase-coherent timing solution (discussed in Section~\ref{timing}). Bottom panel: time residuals with respect to the timing solution.}
\end{figure}

\begin{figure}
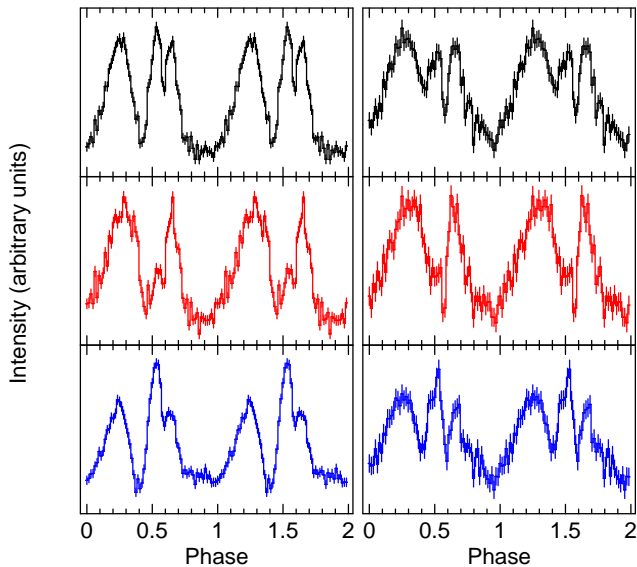

\resizebox{\hsize}{!}{\includegraphics[angle=-90]{pre_norm.eps}\includegraphics[angle=-90]{post_norm.eps}}
\caption{\label{xte_efold} 64-bin \xte\ pulse profiles before (left panel) and after (right panel) the break (see Fig.~\ref{xte_decay} and Section~\ref{rxtespec}). The colours correspond to different energy bands: total (2--10 keV, top, in black), soft (2--4 keV, middle, in red), and hard (4--10 keV, bottom, in blue).}
\end{figure}

\begin{figure}
\resizebox{\hsize}{!}{\includegraphics[angle=-90]{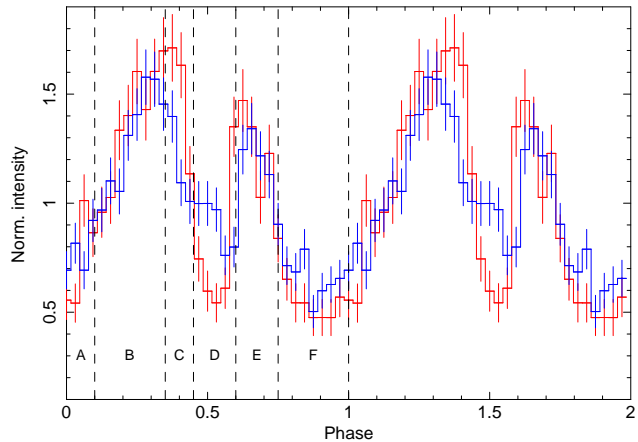}}
\caption{\label{swift_efold} 32-bin \swift/XRT pulse profiles (background-subtracted) of \src\ in the 1.5--2.5 keV (red) and 2.5--10 keV (blue) energy ranges. The vertical dashed lines indicate the phase intervals used for the phase-resolved spectroscopy (see Section~\ref{pps}).}
\end{figure}

Both \xte/PCA and \swift/XRT data were used to study the timing properties of \src. We checked all observations for the presence of bursts by a careful inspection of the light curves binned with different time resolutions, but none was found. We then started the analysis of the coherent pulsations by inferring an accurate period measurement. In order to do this, we folded in the 2--10 keV band the PCA (binned at 10 days) and XRT data at the period reported by \citet*{gogus_atel2076_09} and studied the phase evolution within the observation by means of a phase-fitting technique (details on this technique are given in \citealt{dallosso03}). Given the intrinsic variability of the pulse shape as a function of time, we did not make use of a pulse template to cross-correlate with. We inferred the phase of the modulation by fitting the average pulse shape of each observation with a number of sinusoids, the exact number of which is variable and determined by requesting that the addition of a further (higher) harmonic is not statistically significant (by means of an F-test) with respect to the null hypothesis.\\
\indent The result of this analysis is reported in Fig.~\ref{residual}. The fit of the pulse phases with a linear component results in a reduced $\chi^2_\nu= 1.2$ for 12 dof. The resulting best-fit period is $P=9.078\,388\,4(2)$ s (epoch 54993.0 MJD). This timing solution implies a root mean square (rms) variability of only \mbox{0.03 s}. The inclusion of a quadratic term is not statistically significant and we derive a 3$\sigma$ upper limit on the first period derivative of $|\dot{P}|<1.1\times10^{-13}$ s s$^{-1}$.\\
%Finally, we inferred a 3$\sigma$ upper limit on the magnetic field strength $B_d < 1.1 \times 10^{13}$ Gauss, below the critical value B$_{QED}$=4.4$\times 10^{13}$ Gauss.
%The \xte\ and \swift\ datasets were combined and for each observation, the relative phase of pulsations (epoch: MJD 54993.0) was determined. These phases were then fitted with a linear function, giving a best-fit period of $P= 9.078\,388\,4(2)$ s ($\chi_{\nu}^2=1.3$ for 11 dof). The inclusion of an additional high-order polynomial was not statistically significant. According to our analysis, the 3$\sigma$ upper limit on the period derivative of \src\ is $|\dot{P}|< 1.1 \times 10^{-13}$ s s$^{-1}$.\\ 
\indent The light curve of \src\ in the 2--10 keV band folded at our best period is shown in Fig.~\ref{xte_efold}. 
The pulse morphology is complex, showing two asymmetric peaks per cycle, the second of which exhibits  two subpeaks. 
Because of the problematic background subtraction in the \xte/PCA event-mode data and the possible contamination discussed in Section~\ref{rxtespec}, here we will avoid computing pulsed fractions and hardness ratio, which are meaningful only for background-subtracted data. It is nevertheless evident that the pulse shape changes as a function of energy; one of the subpeaks is in fact prominent in the hard (4--10 keV) band and nearly absent in the soft (2--4 keV) band. 
%, while the rms pulsed fractions are ($4.8\pm0.1$)\% and ($7.5\pm0.2$)\% in the hard and soft bands, respectively. \textbf{[not bkg-subtracted!]}
%\begin{figure}
%\resizebox{\hsize}{!}{\includegraphics[angle=-90]{rxte_efold.jpg}}
%\caption{\label{xte_efold} 64-bin \xte\ pulse profiles (from the whole dataset) of \src\ in different energy bands: total (2--10 keV, in black), soft (2--4 keV, in red), and hard (4--10 keV, in blue).}
%\end{figure}
Furthermore, as already observed by \citet{kuiper09}, the pulse profile evolves with time. This is apparent in Fig.~\ref{xte_efold}, where we show the profiles over the initial phase of the outburst and after the break at about 19 days in the decay curve (see Fig.~\ref{xte_decay} and Section~\ref{rxtespec}).\\
%In the last portion of the light curve, the observed pulsed fractions \textbf{[not bkg-subtracted!]} change from ($6.7\pm0.1$)\% to ($3.0\pm0.1$)\%, from ($10.2\pm0.1$)\% to ($5.4\pm0.2$)\%, and from ($7.5\pm0.1$)\% to $(3.0\pm0.1)\%$ in the 2--10 keV, 2--4 keV, and 4--10 keV the energy bands, respectively.\\
\indent \swift/XRT observations have lower statistics and shorter time coverage, and in PC mode the frame time (2.5 s) does not allow to resolve the pulse fine structures. However, the XRT data have the advantage of providing safe background subtraction and thus the measure of the pulsed fractions. To increase the statistics, we combined the observations as in Section~\ref{swiftspec}. The rms pulsed fractions\footnote{We define the pulsed fraction as \mbox{$\frac{1}{\bar{y}}\big[\frac{1}{n}\sum_{i=1}^{n}(y_i-\bar{y})^2\big]^{1/2}$}, where $n$ is the number of phase bins per cycle, $y_i$ is the number of counts in the $i$-th phase bin, and $\bar{y}$ is the mean number of counts in the cycle.} in the 0.7--10 keV band in the four segments in chronological order are $(23\pm2)\%$, $(34\pm1)\%$, $(26\pm4)\%$, and $(39\pm6)\%$. Apart from an overall increasing trend of the pulsed fraction, which is only marginally significant (i.e. $<$3$\sigma$), the most substantial difference is between the values of the first two segments. We caution, however, that the poorer detail resolution of the PC data, with respect to the WT data of the segment 004--007, can result in a smearing of the pulse profile (and hence in a reduction in the measured pulse fraction) making it difficult to compare the values.\\
\indent The \swift\ folded light curves confirm the variability of the pulse profile with energy; in particular, this is clearly visible in the high resolution WT data shown in Fig.~\ref{swift_efold}, where the pulsed fraction is $(42\pm4)\%$ in the soft (1.5--2.5 keV) band and $(29\pm3)\%$ in the hard (2.5--10 keV) band. To assess the statistical significance of the pulse shape variation, we compared the data using a two-sided Kolmogorov--Smirnov test (see \citealt{press92} and references therein). The result shows that the difference between the soft and hard energy ranges is highly significant: the probability that the two profiles come from the same underlying distribution is in fact $\sim$$1.2\times10^{-4}$.

\section{Phase-resolved spectroscopy}\label{pps}
\begin{figure}
\resizebox{\hsize}{!}{\includegraphics[angle=-90]{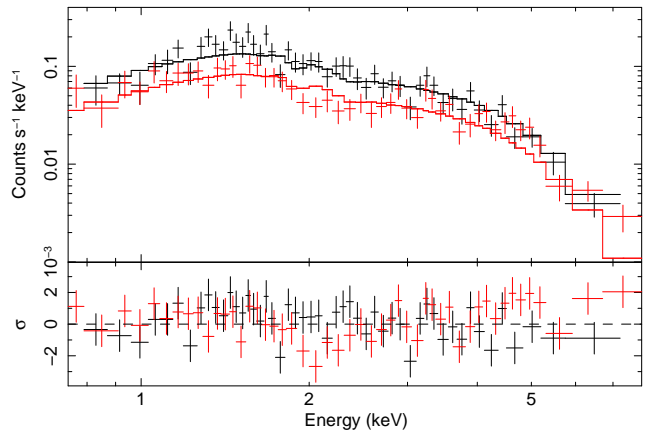}}
\caption{\label{speccd} Phase-resolved spectroscopy of the \swift/XRT observations in WT mode. Top panel: data from the phase intervals C (see Fig.~\ref{swift_efold}), in black, and D, in red. We also plotted (solid lines) the best-fitting blackbody models (with the temperatures forced to a common value, see Section~\ref{pps}). Bottom panel: residuals from the models in units of standard deviations.}
\end{figure}

To study in detail the spectral variations of \src\ with its pulsation phase, we extracted six spectra from the \swift/XRT data in WT mode (obs. 004--007) according to the phase intervals indicated in Fig.~\ref{swift_efold}. We could not use the \swift/XRT data in PC mode because the time resolution of this operating mode is larger than the width of the chosen phase intervals. On the other hand, the \xte/PCA data are not suited either for this kind of analysis due to the large systematic uncertainties affecting the background subtraction and calibration of the PCA data with sufficient time resolution.\\
\indent A simultaneous fit of these spectra with an absorbed blackbody model with only the normalisation allowed to vary gives an acceptable fit ($\chi^2_{\nu}=1.15$ for 365 dof), with spectral parameters consistent with the phase-averaged spectrum. Allowing also the blackbody temperature to vary, a slightly better fit is obtained ($\chi^2_{\nu}=1.09$ for 360 dof), with the highest temperatures in phase D and F ($kT\sim1.04$ keV) and the lowest one during phase C ($kT\sim0.87$ keV).\\
\indent We then concentrated on the two phase intervals with the largest differences, that are also apparent in the \xte\ and \swift\ pulse profiles (Figures~\ref{xte_efold}~and~\ref{swift_efold}): phases C and D. A simultaneous fit of the two spectra with a single blackbody with the same temperature gives $\chi^2_{\nu}=1.26$ for 96 dof, with clearly structured residuals (see Fig.~\ref{speccd}). A slight improvement in the fit is obtained allowing for a different blackbody temperature in the two spectra ($\chi^2_{\nu}=1.12$ for 95 dof), that are $kT= 0.90^{+0.02}_{-0.03}$ keV for phase C and $kT=1.07^{+0.02}_{-0.04}$ keV for phase D. The addition of a second blackbody gives an even better fit ($\chi^2_{\nu}=0.97$ for 91 dof), with the following best-fit parameters ($N_{\rm H}=(4\pm1)\times10^{21}$ cm$^{-2}$ for both spectra and assuming a distance of 5 kpc): $kT_1=0.31^{+0.16}_{-0.06}$ keV, $R_1= 3.4^{+1.2}_{-0.3}$ km, $kT_2=0.90\pm0.02$ keV, $R_2=0.79^{+0.03}_{-0.06}$ km for phase C and $kT_1=0.23\pm0.02$ keV, $R_1=6\pm2$ km, $kT_2=1.14^{+0.06}_{-0.04}$ keV, $R_2=0.43\pm0.04$ km for phase D. A similarly good fit ($\chi^2_{\nu}=0.99$ for 94 dof) but with a reduced number of free parameters is obtained by simultaneously fitting the two spectra with the same single absorbed blackbody model ($N_{\rm H}=(7\pm5)\times10^{20}$ cm$^{-2}$, $kT=0.88\pm0.03$ keV and $R=0.83\pm0.05$ km) and a broad Gaussian absorption line ($E=2.0\pm0.1$ keV, $\sigma_E=1.0\pm0.1$ keV and equivalent width of 1.1 keV) only in the spectrum corresponding to phase D.

\section{Optical observations with the Gran Telescopio Canarias}\label{grantecan}
\begin{figure}
\resizebox{\hsize}{!}{\includegraphics[angle=0]{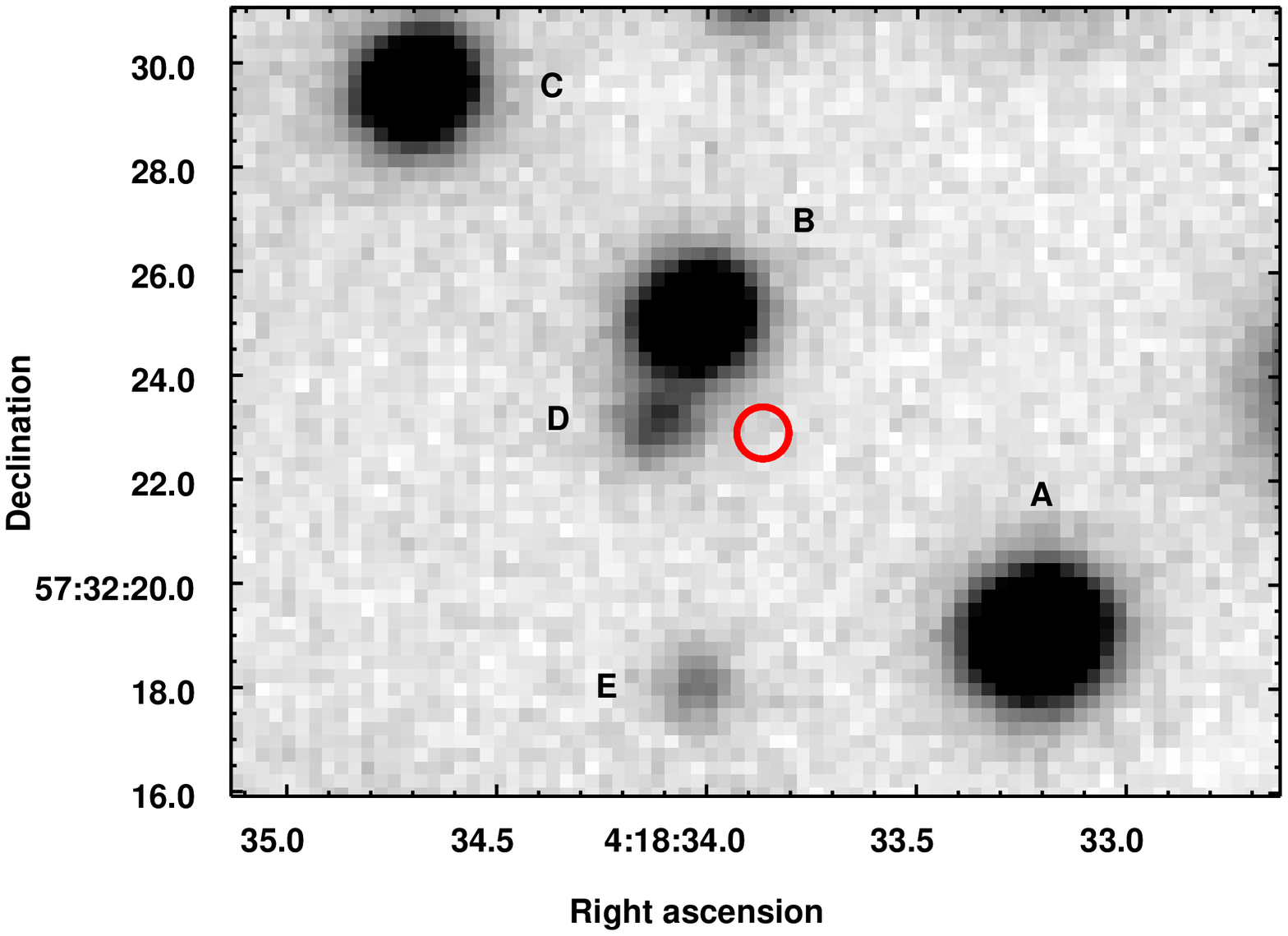}}
\caption{\label{gtc} $20\times15$ arcsec$^2$ section of the $i'$-band image of the field of \src\ as observed by the OSIRIS camera of the GTC. North is up, east to the left. The radius ($0\farcs5$) of the error circle of \src\ (in red) accounts for the $0\farcs35$ uncertainty on the \cxo\ position and for the accuracy of our astrometry calibration (see Section~\ref{grantecan}). Five objects are detected in the field (labeled with capital letters); the observed apparent magnitudes of the brightest sources are: $i'=19.536\pm0.004$ (A), $i'=20.117\pm0.007$ (B), and $i'=20.290\pm0.007$ (C). The two sources (B and D) visible northeast of the \cxo\ error circle are those detected in the $K_{\rm s}$ band by \citet{wachter09}.}
\end{figure}

We observed the field of \src\ through the Sloan $i'$-band filter (central wavelength $\lambda=7800$ \AA) on 2009 September 15 (see Fig.~\ref{xte_decay}) with the Optical System for Imaging and low Resolution Integrated Spectroscopy (OSIRIS; \citealt*{rasilla08}) camera mounted at the 10.4-m Gran Telescopio Canarias (GTC) at the Observatorio del Roque de los Muchachos in the La Palma island (Spain). OSIRIS is a two-chip CCD detector covering the wavelength range between 3650 \AA\ and 10500 \AA, with a nominal fov of $7.8\times 8.5$ arcmin$^2$ which is actually decreased to $7\times7$ arcmin$^2$ due to the vignetting of chip \#1. The unbinned pixel size of the CCD is $0\farcs125$.\\
\indent We took 25 dithered exposures of $\sim$100 s each, for a total integration time of $\sim$2500 s, under dark time and with average seeing conditions of $\sim$$1\farcs3$ (air mass 1.3--1.5). \src\ was positioned in chip \#1, slightly offset from the CCD centre to avoid occultations from the vignetted regions during the dithered exposure sequence. To increase the signal-to-noise ratio, the exposures were taken with a $2\times2$ binning. Bias and dark frames were taken as daily calibrations. In particular, dark frames were taken to correct for the anomalously high dark current affecting the CCD exposures and caused by technical problems. Due to bad weather conditions, no sky flats were taken on 2009 September 15 and those taken on 2009 September 5 were used instead. Data reduction (bias and dark subtraction, flat-field correction) was performed using standard tools in the \textsc{iraf} package\footnote{See http://iraf.noao.edu/.} \textsc{ccdred} after trimming each frame for the vignetted regions. Single dithered exposures were stacked and averaged using the task \textsc{drizzle} which also performs the cosmic ray filtering. Photometry calibration was performed using observations of the standard star field PG\,1528+062 \citep{landolt83}. The astrometry of the science frames was calibrated against the positions and coordinates of Two-Micron All-Sky Survey (2MASS; \citealt{skrutskie06short}) stars, resulting in an overall accuracy of $\sim$$0\farcs1$ in both Right Ascension and Declination.\\
\indent The computed \src\ position on the OSIRIS image is shown in Fig.~\ref{gtc}. We did not detect any source within the 95\%-confidence \cxo\ error circle (radius of $0\farcs35$) of \src\ \citep{woods09}. In particular, we did not find evidence for the possible detection of the infrared source ($K_{\rm s} = 21.6 \pm 1.3$ mag) tentatively seen by \citet{wachter09} southwest of the \cxo\ error circle on 2009 August 2, using the Wide-field InfraRed Camera (WIRC) of the Hale 5.1-m telescope at Palomar Observatory (California, USA). We set a $3\sigma$ limiting magnitude of $i'> 25.1$ on the optical brightness of \src\ \citep{mrt09short}, which is the deepest obtained so far in the optical band for this source. The previous optical upper limit, obtained from observations performed with the Auxiliary-port Camera (ACAM) Imager at the 4.2-m William Herschel Telescope (WHT) on La Palma one month earlier, on 2009 August 16, was in fact $r>24$ mag \citep{ratti09}. After de-reddening our measured upper limit using the value of the hydrogen column density derived from the X-ray spectral fits\footnote{We used the relation between the \nh\ and the interstellar extinction $A_{\rm{V}}$ of \citet{predehl95}, as well as the relations between the extinctions at different wavelengths of \citet{fitzpatrick99}.} (Table~\ref{xrt-spec}), and using the unabsorbed 1--10 keV X-ray flux measured by \swift/XRT (observations 008--009, which were carried out at an epoch close to the GTC observations), we infer an X-ray-to-optical flux ratio of $F_{\mathrm{X}}/F_{i'} > 10\,000$ (for a bandwidth $\Delta\lambda=1500$ \AA), in line with the ratio observed for other magnetars (e.g. \citealt{mignani09}).\\
\indent Similarly, we constrained the X-ray-to-near-infrared flux ratio using as a reference the Hale telescope observations, for which we assumed a non-detection of \src\ down to $K_{\rm s}=22.9$ mag. Since no \swift/XRT observations were obtained quasi-simultaneously to these observations, we assumed the 1--10 keV X-ray flux obtained from the monitoring \xte/PCA observations. After de-reddening, we inferred $F_{\mathrm{X}}/F_{K_{\mathrm{s}}}>50000$. The X-ray-to-optical flux ratio upper limit derived from WHT observations is $F_{\mathrm{X}}/F_{r}>3000$. The uncertainties on these ratios, mainly due to the uncertainty in the \xte/PCA flux determination, are of the order of $\approx$$20\%$.

\section{Pre-outburst X-ray observations}\label{rosat}
There are two \emph{R\"ontgen Satellite} (\rst) All-Sky Survey \citep{voges99short} datasets containing the region of \src\ (sequences rs930707n00 and rs930708n00); the data were collected in the period between 1990 August 8 and 1991 February 18. The field of \src\ received each time roughly 530 s of exposure with the Position Sensitive Proportional Counter (PSPC; \citealt{pfeffermann87short}).\\
\indent \src\ was not detected in these data and the upper limit on the count rate, computed as the 3$\sigma$ noise level due to the background at the SGR location, is about 0.05 counts s$^{-1}$ in the 0.4--2.4 keV energy range. Using the \textsc{webpimms}\footnote{See http://heasarc.gsfc.nasa.gov/Tools/w3pimms.html.} (v3.9j) tool and assuming an absorbed blackbody with $kT=0.8$ keV and $N_{\rm H}=1.1\times10^{21}$ cm$^{-2}$ (see Table~\ref{xrt-spec}), the limit translates into an absorbed 1--10 keV (2--10 keV) flux of $\approx$$3\times10^{-12}$ \flux\ ($\approx$$2\times10^{-12}$ \flux).

\section{Discussion and conclusions}\label{disc}
A short period of  bursting activity recently led to the discovery of the soft gamma-ray repeater \src\ \citep{vanderhorst09short,vanderhorst10short}. In the first days after the onset of the outburst, the observed flux was of a few $10^{-11}$ \flux, and it decreased by a factor $\sim$10 during the following five months. It is unknown whether the flux of \src\ has already decayed to (or dropped below) the pre-outburst level, since the archival (\rst) data of the field provide only an upper limit of $\approx$$2\times10^{-12}$ \flux\ and magnetars often display flux variations over an order of magnitude or more (see e.g. \citealt{gh07,icd07,eiz08short}).\\ 
\indent The X-ray data presented here allowed us to measure with accuracy the spin period of \src\ and to reassess the results reported in \citet{israel09}, where a (marginally significant) period derivative of $\dot{P} = 3.0(9)\times 10^{-13}$ s s$^{-1}$ was proposed. Now, on a longer baseline, we only infer a phase-coherent 3$\sigma$ limit on the period derivative of $|\dot{P}|< 1.1 \times 10^{-13}$ s s$^{-1}$ (valid over the range MJD 54993--55144; see \citealt{kuiper09} and \citealt*{wgk09} for previous upper limits). The corresponding upper limits on the surface dipole magnetic field strength (inferred within the usual vacuum dipole framework; see e.g. \citealt{lorimer04}) and spin-down luminosity are $B \approx 3.2\times10^{19}(P\dot{P})^{1/2}< 3 \times 10^{13}$ G and $\dot{E}=4\pi^2 I \dot{P}P^{-3}<6\times10^{30}$ \lum, respectively, where $I\simeq10^{45}$ g cm$^2$ is the moment of inertia of a neutron star. The properties inferred from the period derivative should be taken with additional caution for magnetars, since their spin-down rates can be highly variable. Nevertheless, the present limits on $B$ and $\dot{E}$ for \src\ are the lowest values among magnetars (see e.g. Figure~3 of \citealt{ebp09short}), 1E\,2259+586 being the closest, with $B\simeq6\times 10^{13}$ G and $\dot{E}\simeq 5.6\times10^{31}$ \lum\ \citep{gk02}.\\ 
\indent The distance of \src\ is not known yet; however, considering the structure of the Galaxy proposed by \citet*{hou09}, the Galactic coordinates of the SGR ($l= 147\fdg98$ and $b =5\fdg12$) suggest that it is located within a distance smaller than $\approx$5 kpc. \citet{vanderhorst10short} assume a distance of $\sim$2 kpc, in the hypothesis that \src\ is situated in the Perseus arm of the Galaxy. Such a relatively small distance is consistent with the measure of the photoabsorption derived from the X-ray spectral analysis ($N_{\rm H}\simeq1.1\times10^{21}$ cm$^{-2}$, see Table~\ref{xrt-spec}). This value is in fact significantly smaller than the total Galactic interstellar hydrogen alone the line of sight which can be estimated from the Leiden/Argentine/Bonn (LAB) survey of H\,I \citep{kalberla05} and the CO survey of \citet*{dame01} for the value of molecular H$_2$ as $N_{\mathrm{H}}=N_{\mathrm{H\,I}}+2N_{\mathrm{H_2}}=(4.30+2\times0.67)\times10^{21}$ cm$^{-2}\simeq5.6\times10^{21}$ cm$^{-2}$.\\ 
\indent Conservatively assuming a distance of 5 kpc, the 1--10 keV X-ray luminosity measured with \swift\ varied during the outburst from $L_{\rm{X}}\simeq4\times10^{34}$ \lum\ to $6\times10^{33}$ \lum. The minimum observed ratio between the luminosity and the spin-down energy loss is $L_{\rm{X}}/\dot{E}> 10^3d_5^2$, where $d_{\rm{N}}$ indicates the distance in units of N kpc, a value which, when compared to those of other magnetar sources, fully qualifies \src\ as a member of the magnetar class.\\ 
%\indent The upper limit on the period derivative corresponds, assuming a pure magneto-polar spin down, to a characteristic age $\tau_c = P/(2\dot{P})> 1.1$ Myears, which potentially would make \src\ the oldest magnetar discovered so far. Interestingly enough, its location on the $P$--$\dot{P}$ diagram (e.g. figure 4 of \citealt{mclaughlin09short}) falls quit close to the region of the X-ray dim isolated neutron stars (XDINSs; see e.g. \citealt{kaplan08}), often proposed to be possibly evolved magnetars on the basis of their comparable spin periods and relatively large dipole magnetic field. \src\ might thus be a case of an old magnetar, or young XDINS, ideally linking the two classes of isolated neutron stars.
%\\ \indent We searched the radio supernova remnant (SNR) database by \citet{green09} and the master supernova remnant catalogues available
%through \emph{Simbad}\footnote{See http://simbad.u-strasbg.fr/simbad/.}
%but we did not find any entry within a very generous radius of $2^{\circ}$
%from \src. The lack of an optical/radio SNR, which could be still
%detectable at the \src\ distance and for the relatively low estimated
%absorption along the line of sight, might suggest a relatively old age for
%the magnetar. Tantalising as it might be, we note that the SNR
%non-detection can be obviously due to an observational bias like, e.g. the
%lack of SNR radio surveys at these relatively high Galactic latitude.
%Deep, targeted observations in the radio band are required to search for a
%possibly associated SNR. \\
\indent The upper limit on the period derivative corresponds, in the standard magneto-dipole braking model, to a characteristic age $\tau_c = P/(2\dot{P})> 1.1$ million years, the highest (so far) among magnetars. The location of \src\ on the $P$--$\dot{P}$ diagram (see e.g. Figure~4 of \citealt{mclaughlin09short} for an updated diagram) falls quite close to the region of the X-ray dim isolated neutron stars (XDINSs; see \citealt{turolla09} for a review), often proposed to be evolved magnetars on the basis of their comparable spin periods and relatively large dipole magnetic fields. Although age estimates based on the magneto-dipole model are highly uncertain\footnote{An independent age estimate could come from the association with a supernova remnant. However, there are no known remnants within $2^{\circ}$ from \src.} it is intriguing to think \src\ as an `old magnetar' (or a young XDINS), ideally linking the two classes of isolated neutron stars.\\ 
\indent We performed the spectral analysis of the \swift\ observations of \src\ by assuming different phenomenological models (combinations of blackbodies and power laws) or more complex ones, based on resonant cyclotron scattering of seed surface photons by magnetospheric currents (1D and NTZ models). In all cases, we found that \swift\ data are compatible with a scenario in which the flux decrease is mainly due to the cooling and shrinking of a hot emitting region on the neutron-star surface, which may have been suddenly heated following the bursting activity. This is in agreement with what is observed in the post-outburst evolution of other transient magnetars (e.g. \citealt{gotthelf05}). SGR\,0501+4516 is the only other repeater for which the outburst decay has been monitored with detailed, quasi-continuous observations (\xmm \ and \swift)  over a time span ($\sim$160 days) comparable to the present one \citep{rea09}. In that case, however, the flux decline was exponential with a timescale $\sim$24 days and no evidence for a power-law decay. The first \swift\ observation was performed only 33 days after the onset of the bursting phase and by that time one may expect that major deviations from a dipolar magnetosphere have died away (see e.g. \citealt{eiz08short}), leaving only a cooling hot spot at the neutron-star surface as the relic of the outburst. Alternatively, the twisted portion of the magnetosphere could have been quite small, as suggested by \citet{beloborodov09} in the case of the AXP XTE\,J1810--197, making it difficult for thermal photons to resonantly scatter onto magnetospheric currents.\\ 
\indent \xte\ pulse profiles (and, to a lesser extent because of the lower count statistics, also \swift\ ones) exhibit a complex pattern, with an overall double-peaked shape. If we assume that emission mainly comes from the star surface with little contribution from the magnetosphere (see the discussion above) and it is essentially isotropic,\footnote{We warn that this is an oversimplification since thermal emission from a magnetised neutron star, either from an atmosphere or a condensed surface, is not isotropic (e.g. \citealt{zavlin95}; \citealt*{turolla04}).} double-peaked light curves arise when the angles $\xi$, between the line connecting the two (antipodal) emitting spots and the rotation axis, and $\chi$, between the line-of-sight and the rotation axis, are such that $\xi + \chi \ga 120^\circ$ for a radius $R_{\mathrm{NS}}\sim 3 R_S$ (here $R_S$ is the Schwarzschild radius; e.g. \citealt{beloborodov02,poutanen06}). This suggests that \src\ is a nearly orthogonal rotator seen at a large inclination angle. Pulse profiles are both energy- and time-dependent, with evidence for a hard ($>$4 keV) component concentrated near the second light maximum (phase $\sim$0.6), which becomes less prominent as the flux declines (see Fig.~\ref{xte_efold}). This, again, points to a picture in which only a limited portion of the star surface has been heated as a consequence of the event which produced the outburst. The near coincidence in phase of the (second) soft and hard peaks, together with the decrease of the latter at later times, may indicate that heating involved only one of the two spots, e.g. producing a hotter region inside (or close to) one of the warm polar caps.\\ 
\indent Phase-resolved spectroscopy of the (combined) \swift\ WT data shows that emission near the phase intervals of maximum softness/hardness is described well by the superposition of two blackbodies with different radii and temperatures in each interval. The radii of the two cooler blackbodies are not much different and, although somewhat too small ($\approx 4$--6 km), might be associated with the neutron-star radius, while their temperatures are quite close and consistent with being the same within the errors. A possibility, then, is that emission comes from two spots at different temperature, while the rest of the surface is at a lower temperature.
%We calculated the pulse profiles in the three \xte\ energy bands using the radii and temperatures of the two spots as derived from the spectral fits (see Section~\ref{pps}; $R_{\mathrm{NS}}=13$ km, $kT_{\mathrm{cold}}= 0.25$ keV) and assuming an orthogonal rotator seen edge-on ($\xi=\chi=90^\circ$). We find that results are in qualitative agreement with the observed lightcurves after the break, although we did not attempt any systematic exploration of the parameter space. We repeated the same calculation for the other two options implied by the
Phase-resolved spectral fits suggest two other options: two emitting regions, each comprising a hotter inner cap and a surrounding warmer corona, and two equal spots, one of which emits a blackbody plus a Gaussian absorption feature. In both these scenarios the rest of the surface is likely to be rather colder than the lower temperature obtained from the blackbody fit. A preliminary calculation indicates that all these configurations can produce pulse profiles which are in general agreement with the observed ones.\\
%$kT_{\mathrm{cold}}=0.05$ keV). The agreement is somewhat worse, although this by no means can be taken as conclusive evidence in favour of the first hypothesis.
\indent After the outburst, the flux of \src\ follows a broken power-law decay, $F\propto t^\alpha$, with a break at $\sim$19 days and $\alpha_1\sim -0.3$, $\alpha_2\sim -1.2$ (for the fit to the combined \xte\ and \swift\ data; see Section~\ref{rxtespec}). The post-outburst cooling of magnetars has been modelled in terms of a sudden heat deposition in the neutron-star crustal layers, due to the release of energy stored in the toroidal component of the internal magnetic field (\citealt*{let02}; see also \citealt{kouveliotou03short}). In this picture the flux decays as $F\propto t^{-n/3}$, with $n\sim 2$--3. While the predicted index might account for the flux evolution in \src\ after the break, it is inconsistent with what is observed before the break, and, even more important, the model does not predict any change of slope during the decay. On the other hand, it should be stressed that the calculation of \citet*{let02} was mainly aimed at explaining the flux decay in SGR~1900+14 after its giant flare and consequently assumes a magnetic field $B\sim 10^{15}\, {\rm G}$. According to the limit on $\dot P$ presented in Section~\ref{timing}, the magnetic field of \src\ is likely to be two orders of magnitude lower, so the model may not be directly applicable in this case.\\ 
\indent A different possibility is that the surface layers of the star are heated by returning currents that flow along the closed field lines of a twisted magnetosphere \citep*{tlk02,beloborodov07,beloborodov09}. The twist is likely to be localised into a small bundle of current-carrying field lines (the $j$-bundle) close to the magnetic pole and must decay in order to supply its own supporting currents. As the magnetosphere untwists, the angular extent of the $j$-bundle decreases, the luminosity decays and the area hit by the currents shrinks \citep{beloborodov09}. The details of the evolution depend on the initial distribution of the twist inside the $j$-bundle and no general law for the luminosity decay exists. If the twisted region is small, the emergent spectrum is dominated by emission from the heated cap and it is thermal because photons do not have many chances to scatter. The flux is then $F\propto R^2T^4$ where $R$ and $T$ are the radius and temperature of the cap. The decay of the temperature (as derived from the single blackbody fits to \xte\ and \swift\ data) is reasonably well described by a single power-law, $T\propto t^{-0.05}$ (see Fig.~\ref{xte_decay}), over the entire time span, without the need to introduce a break. If the temperature variation is entirely due to the cooling of a cap, the emitting area decreases as $R^2\propto t^{-0.08}$ and $R^2\propto t^{-0.97}$ before and after the break, respectively. If the single blackbody model mostly traces the emission from the hot region, as it seems reasonable also from the comparison of the fit parameters with those of the hotter component in the two blackbody model (see Table~\ref{xrt-spec} and Section~\ref{pps}), it follows that, after the break, the radius of the cap scales as $t^{-0.41}$, close enough to the dependence derived above ($R\propto t^{-0.48}$). A better estimate of the cap radius could come from the two blackbody model, but the limited statistics allows its application only to the combined \swift\ observations, taken at different epochs after the break. Assuming as a rough estimate that the hottest blackbody radius $R_2\sim 0.43$ km can be associated with $t\sim 35$ days, around which five observations cluster, the source luminosity is $L\sim 4\times 10^{34} d_5^2$ \lum. This value is too high to be explained by the rate of Ohmic dissipation, $L\approx 10^{36} (B/10^{14}\, {\rm G})(R_\mathrm{NS}/10^6\, {\rm cm})\psi (V/10^9\, {\rm V}) \sin^4\theta_*$ \lum (\citealt{beloborodov09}; here $\psi\la 1\, {\rm rad}$ is the twist angle and $V$ is the discharge voltage) if the magnetic field is $\sim 10^{13}$ G and the angular size of the twist is $\sin\theta_*\sim R_2/R_{\mathrm{NS}}\sim 0.03$. The problem is however eased for a larger cap area, as in the case in which the cap contains a small hot region surrounded by a warm corona.\\
%We have computed the pulse profiles before the break (we take as a reference $t = 10$ days) estimating the radii and temperature from the scaling relations derived above. Both for two spots at uniform $T$ superimposed to a cooler surface at 0.25 keV and for two caps with an inner hotter zone and a warmer corona, the hard peak becomes more pronounced with respect to what occurs after the break, in general agreement with the observed pulse profiles. 
%\indent Finally, we note that, although sampling different parts of the X-ray light curve, comparing the $F_{\mathrm{X}}/F_{K_{\mathrm{s}}}$ and $F_{\mathrm{X}}/F_{i'}$ upper limits does not provide useful information to constrain the flux decay in the optical/IR, given the lack of a template spectral model of the magnetar emission at these wavelengths (e.g. \citealt{mignani09}). On the other hand, the comparison with similar flux ratios measured for other sources is hampered by the extreme variability of these sources, with such flux ratios varying erratically along the X-ray light curve decay (e.g. \citealt{testa08short}) and spanning at least one order of magnitude (see, e.g. discussion in \citealt{mignani09}). 
\indent Currently, several AXPs and SGRs have been firmly identified also at optical and/or infrared wavelengths (see \citealt{mignani09} and references therein). However, there are no detailed predictions yet about the X-ray-to-optical (or infrared) flux ratio that a magnetar is expected to have. Our deep GTC $i'$ observation of \src\ points to an X-ray-to-optical flux ratio exceeding $10^4$, a value not unheard of among magnetars (e.g. \citealt{mignani09}). The GTC image adds a new piece to the still sparse -- but growing -- database of optical/infrared observations of magnetars exploring a range of different states of high energy activity. Enlarging such a database is indeed a farsighted effort since it will be the only way to decipher the nature of magnetar optical/infrared emission and of its correlation with the high energy emission.

\section*{Acknowledgments}
This research is based on observations with the NASA/UK/ASI mission \swift\ and the NASA mission \xte. We thank the \swift\ PI, Neil Gehrels, the \swift\ duty scientists and science planners for making our \swift\ target of opportunity observations possible. The \xte\ and \swift\ data were obtained through the High Energy Astrophysics Science Archive Research Center (HEASARC) Online Service, provided by the NASA/Goddard Space Flight Center. We thank Jerome Rodriguez and Isabel Caballero for the help with the \xte\ data. This work also made use of observations carried out with the Gran Telescopio Canarias (GTC), installed in the Spanish Observatorio del Roque de los Muchachos of the Instituto de Astrof\'isica de Canarias, in the island of La Palma. We wish to thank Rene Rutten, Riccardo Scarpa, and the whole GTC staff for their technical support and prompt reaction. The Italian authors acknowledge the partial support from ASI (ASI/INAF contracts I/011/07/0, I/010/06/0, I/088/06/0, and AAE~TH-058). DG, SC, and FM acknowledge the CNES for financial funding. RPM and SZ acknowledge support from STFC. NR is supported by a Ram\'on~y~Cajal fellowship.

\bibliographystyle{mn2e}
\bibliography{biblio}

\bsp

\label{lastpage}

\end{document}